%

%
\input harvmac.tex
\input amssym.def
\def\CA{{\cal A}}

\def\CE{{\cal E}}

\def\CT{{\cal T}}

\def\CN{{\cal N}}
\def\CO{{\cal O}}
\def\CP{{\cal P}}

\def\CW{{\cal W}}

\def\CY{{\cal Y}}

\def\hl{{\hat \l}}
\def\t{\theta}

\font\cmss=cmss10 \font\cmsss=cmss10 at 7pt
\def\IR{\relax{\rm I\kern-.18em R}}
\def\inbar{\vrule height1.5ex width.4pt depth0pt}
\def\IC{\relax\,\hbox{$\inbar\kern-.3em{\rm C}$}}
\def\IN{\relax{\rm I\kern-.18em N}}
\def\IF{\relax{\rm I\kern-.18em F}}
\def\IP{\relax{\rm I\kern-.18em P}}
\def\IZ{\relax\ifmmode\mathchoice
{\hbox{\cmss Z\kern-.4em Z}}{\hbox{\cmss Z\kern-.4em Z}}
{\lower.9pt\hbox{\cmsss Z\kern-.4em Z}} {\lower1.2pt\hbox{\cmsss
Z\kern-.4em Z}}\else{\cmss Z\kern-.4em Z}\fi}
\def\sm#1#2{\kern.1em\lower.4ex\hbox{${\scriptstyle #2}$\kern-1em\raise1.6ex
             \hbox{${\scriptstyle #1}$}}}
\def\sfrac#1#2{\textstyle{#1\over #2}}

\def\underrightarrow#1{\vbox{\ialign{##\crcr$\hfil\displaystyle
{#1}\hfil$\crcr\noalign{\kern1pt
\nointerlineskip}$\longrightarrow$\crcr}}}
\def\underleftarrow#1{\vbox{\ialign{##\crcr$\hfil\displaystyle
{#1}\hfil$\crcr\noalign{\kern1pt
\nointerlineskip}$\longleftarrow$\crcr}}}

\def\cob#1{\overline{\CO}(#1)}

\def\hCU{{\hat\CU}}

\def\i{{\rm i}}
\def\b{{\beta}}
\def\bd{{\dot{\beta}}}
\def\a{{\alpha}}
\def\ad{{\dot{\alpha}}}
\def\d{{\delta}}

\def\g{{\gamma}}
\def\gd{{\dot{\gamma}}}
\def\o{{\omega}}

\def\t{{\theta}}

\def\l{{\lambda}}

\font\mtt=cmr9
\font\mt=cmr12

\baselineskip=14pt plus 2pt minus 1pt
%
%

\lref\PopovRB{
      A.~D.~Popov and C.~S\"amann,
      {\it On supertwistors, the Penrose-Ward transform and $\CN=4$ super 
      Yang-Mills theory},
      hep-th/0405123.
}

\nref\WittenNN{
      E.~Witten,
      {\it Perturbative gauge theory as a string theory in twistor space},
      hep-th/0312171.
}

\nref\WittenFB{
      E.~Witten,
      {\it Chern-Simons gauge theory as a string theory},
      Prog.\ Math.\  {\bf 133} (1995) 637
      [hep-th/9207094].
}

\nref\BerkovitsHG{
      N.~Berkovits,
      {\it An alternative string theory in twistor space for $\CN = 4$ 
          super-Yang-Mills},
      Phys.\ Rev.\ Lett.\  {\bf 93} (2004) 011601
      [hep-th/0402045];
      {\it Self-dual super-Yang-Mills as a string theory in $(x,\t)$-space},
      JHEP {\bf 0405} (2004) 034
      [hep-th/0403280].
}

\nref\NeitzkePF{
      A.~Neitzke and C.~Vafa,
      {\it $N=2$ strings and the twistorial Calabi-Yau},
      hep-th/0402128;
      N.~Nekrasov, H.~Ooguri and C.~Vafa,
      {\it S-duality and topological strings},
      hep-th/0403167.
}

\nref\AganagicYH{
      M.~Aganagic and C.~Vafa,
      {\it Mirror symmetry and supermanifolds},
      hep-th/0403192;
      S.~P.~Kumar and G.~Policastro,
      {\it Strings in twistor superspace and mirror symmetry},
      hep-th/0405236.
}

\nref\BerkovitsTX{
      N.~Berkovits and L.~Motl,
      {\it Cubic twistorial string field theory},
      JHEP {\bf 0404} (2004) 056
      [hep-th/0403187].
}

\nref\SiegelDJ{
      W.~Siegel,
      {\it Untwisting the twistor superstring},
      hep-th/0404255.
}

\nref\BerkovitsJJ{
      N.~Berkovits and E.~Witten,
      {\it Conformal supergravity in twistor-string theory},
      JHEP {\bf 0408} (2004) 009
      [hep-th/0406051].
}

\nref\LechtenfeldCC{
      O.~Lechtenfeld and A.~D.~Popov,
      {\it Supertwistors and cubic string field theory for open $N=2$ 
      strings}, 
      Phys.\ Lett.\ B {\bf 598} (2004) 113
      [hep-th/0406179].
}

\lref\Manin{
      Yu.~I.~Manin,
      {\it Gauge field theory and complex geometry},
      Springer, Berlin (1988).
}

\lref\LechtenfeldIK{
      O.~Lechtenfeld and A.~D.~Popov,
      {\it Closed N = 2 strings: Picture-changing, hidden symmetries and SDG
      hierarchy},
      Int.\ J.\ Mod.\ Phys.\ A {\bf 15} (2000) 4191
      [hep-th/9912154];
      T.~A.~Ivanova and O.~Lechtenfeld,
      {\it Hidden symmetries of the open N = 2 string},
      Int.\ J.\ Mod.\ Phys.\ A {\bf 16} (2001) 303
      [hep-th/0007049].
}

\nref\NairBQ{
      V.~P.~Nair,
      {\it A current algebra for some gauge theory amplitudes},
      Phys.\ Lett.\ B {\bf 214} (1988) 215.
}

\nref\RoibanVT{
      R.~Roiban, M.~Spradlin and A.~Volovich,
      {\it A googly amplitude from the $B$-model in twistor space},
      JHEP {\bf 0404} (2004) 012
      [hep-th/0402016];
      {\it On the tree-level $S$-matrix of Yang-Mills theory},
      Phys.\ Rev.\ D {\bf 70} (2004) 026009 
      [hep-th/0403190];
      R.~Roiban and A.~Volovich,
      {\it All googly amplitudes from the $B$-model in twistor space},
      hep-th/0402121;
      G.~Georgiou and V.~V.~Khoze,
      {\it Tree amplitudes in gauge theory as scalar MHV diagrams},
      JHEP {\bf 0405} (2004) 070
      [hep-th/0404072].
}

\nref\CachazoKJ{
      F.~Cachazo, P.~Svrcek and E.~Witten,
      {\it MHV vertices and tree amplitudes in gauge theory},
      JHEP {\bf 0409} (2004) 006
      [hep-th/0403047];
      {\it Twistor space structure of one-loop amplitudes in gauge theory},
      hep-th/0406177;
      C.~J.~Zhu,
      {\it The googly amplitudes in gauge theory},
      JHEP {\bf 0404} (2004) 032
      [hep-th/0403115];
      J.~B.~Wu and C.~J.~Zhu,
      {\it MHV vertices and scattering amplitudes in gauge theory},
      JHEP {\bf 0407} (2004) 032
      [hep-th/0406085];
      {\it MHV vertices and fermionic scattering amplitudes in gauge theory 
      with quarks and gluinos},
      hep-th/0406146.
}

\nref\WittenCP{
      E.~Witten,
      {\it Parity invariance for strings in twistor space},
      hep-th/0403199;
      S.~Gukov, L.~Motl and A.~Neitzke,
      {\it Equivalence of twistor prescriptions for super Yang-Mills},
      hep-th/0404085.
}

\nref\GiombiIX{
      S.~Giombi, R.~Ricci, D.~Robles-Llana and D.~Trancanelli,
      {\it A note on twistor gravity amplitudes},
      JHEP {\bf 0407} (2004) 059
      [hep-th/0405086];
      N.~Berkovits and E.~Witten,
      {\it Conformal supergravity in twistor-string theory},
      JHEP {\bf 0408} (2004) 009
      [hep-th/0406051].
}

\lref\PopovPC{
      A.~D.~Popov,
      {\it Self-dual Yang-Mills: Symmetries and moduli space},
      Rev.\ Math.\ Phys.\  {\bf 11} (1999) 1091
      [hep-th/9803183].
}

\lref\PopovFB{
      A.~D.~Popov,
      {\it Holomorphic Chern-Simons-Witten theory: From $2D$ to $4D$ conformal 
      field theories},
      Nucl.\ Phys.\ B {\bf 550} (1999) 585
      [hep-th/9806239];
      T.~A.~Ivanova and A.~D.~Popov,
      {\it Dressing symmetries of holomorphic BF theories},
      J.\ Math.\ Phys.\  {\bf 41} (2000) 2604
      [hep-th/0002120].
}

\lref\SiegelZA{
      W.~Siegel,
      {\it $N=2$ $(4)$ string theory is self-dual $\CN=4$ Yang-Mills theory},
      Phys.\ Rev.\ D {\bf 46} (1992) R3235
      [hep-th/9205075];
      G.~Chalmers and W.~Siegel,
      {\it The self-dual sector of QCD amplitudes},
      Phys.\ Rev.\ D {\bf 54} (1996) 7628
      [hep-th/9606061].
}

\lref\WittenZE{
      E.~Witten,
      {\it Topological quantum field theory},
      Commun.\ Math.\ Phys.\  {\bf 117} (1988) 353.
}

\lref\VafaTF{
      C.~Vafa and E.~Witten,
      {\it A Strong coupling test of $S$-duality},
      Nucl.\ Phys.\ B {\bf 431} (1994) 3
      [hep-th/9408074].
}

\lref\PenroseIN{
      R.~Penrose,
      {\it The twistor program},
      Rept.\ Math.\ Phys.\  {\bf 12} (1977) 65.
}

\newbox\tmpbox\setbox\tmpbox\hbox{\abstractfont }

\Title{ \vbox{
\rightline{\hbox{hep-th/0406224}}
\rightline{\hbox{ITP--UH--20/04}}}}
{\vbox{\centerline{Topological B-Model on Weighted Projective Spaces}
\vskip .2cm
\centerline{and Self-Dual Models in Four Dimensions}}}
\smallskip
\centerline{{\mt Alexander D. Popov $^*$ and Martin Wolf}}
\smallskip
\centerline{\it Institut f\"ur Theoretische Physik}
\centerline{\it Universit\"at Hannover}
\centerline{\it Appelstra{\ss}e 2, 30167 Hannover, Germany}
\smallskip
\centerline{E-mail: \tt popov, wolf@itp.uni-hannover.de}
\vskip 1cm
\centerline{\bf Abstract}
\smallskip
It was recently shown by Witten on the basis of several examples that the 
topological $B$-model whose target space is a Calabi-Yau (CY) supermanifold is 
equivalent to holomorphic Chern-Simons (hCS) theory on the same supermanifold. 
Moreover, 
for the supertwistor space $\IC P^{3|4}$ as target space, it has been
demonstrated that hCS theory on  $\IC P^{3|4}$ is equivalent to self-dual
$\CN=4$ super Yang-Mills (SYM) theory in four dimensions. We consider
as target spaces for the $B$-model the weighted projective spaces 
$W \IC P^{3|2}(1,1,1,1|p,q)$ with two fermionic coordinates of weight $p$ and 
$q$, respectively -- which are CY supermanifolds for $p+q=4$ -- and discuss 
hCS theory on them. By using twistor techniques, we obtain certain field 
theories in four dimensions which are equivalent to hCS theory. These theories 
turn out to be self-dual truncations of $\CN=4$ SYM theory or of its 
twisted (topological) version.

\vfill
\noindent June, 2004
\smallskip\smallskip
\hrule width 5cm
\smallskip\smallskip
{\mtt $^*$ On leave from Laboratory of Theoretical Phyiscs, JINR, Dubna, 
Russia.}

\supereject\footline={\hss\tenrm\folio\hss}

\newsec{Introduction and results}

Recently, Witten has shown \WittenNN\ that there is a one-to-one correspondence
between the moduli spaces of holomorphic Chern-Simons (hCS) theory \WittenFB\
on the Calabi-Yau (CY) supermanifold $\IC P^{3|4}\setminus\IC P^{1|4}$ and of 
self-dual $\CN=4$ super Yang-Mills (SYM) theory on $\IR^4$ with a 
metric of signature $({+}\,{+}\,{+}\,{+})$ or $({+}\,{+}\,{-}\,{-})$. It was 
also shown that hCS theory on 
$\IC P^{3|4}$ is related with the $B$-type open topological string 
(the $B$-model)
having $\IC P^{3|4}$ as target space (for related works see \BerkovitsHG\ -- 
\LechtenfeldCC). Therefore, string theory considerations can help
to clarify properties of quantum $\CN=4$ SYM theory (see e.g.
\NairBQ\ -- \WittenCP).\foot{For properties of gravity amplitudes see
e.g. \GiombiIX.}

Recall that the fermionic coordinates $\eta_i$ ($i=1,\ldots,4$) of the 
supermanifold $\IC P^{3|4}= W\IC P^{3|4}(1,1,1,1|1,1,1,1)$ take values in the 
holomorphic line bundle $\CO(1)$. As the product $\prod_{i=1}^4\,{\rm d}\eta_i$
takes values in $\CO(-4)$ and the holomorphic measure on $\IC P^3$ is $\CO(4)$ 
valued, the full holomorphic measure on $\IC P^{3|4}$ is globally defined and 
nowhere vanishing. Thus, the supertwistor space $\IC P^{3|4}$ is a CY 
supermanifold \WittenNN. In this note, we will focus on the weighted projective
spaces $W\IC P^{3|2}(1,1,1,1|p,q)$ with $(p,q)$ equal to $(1,3)$, $(2,2)$ and 
$(4,0)$. Since $p+q=4$ for these spaces, they are CY supermanifolds due to the
above reasoning. We describe hCS theory on open subsets of these weighted 
projective spaces and derive (equivalent) field theories on $\IR^4$ from it. 
We show that these models are related with different self-dual truncations of 
either $\CN=4$ SYM theory or its twisted (topological) version.

\newsec{Twistor geometry}

{\bf Local coordinates.} 
Our starting point is the complex projective space $\IC P^3$ (twistor space)
with homogeneous coordinates $(\o^\a,\l_\ad)$. These coordinates
are subject to the equivalence relation 
$(\o^\a,\l_\ad)\sim(t\o^\a,t\l_\ad)$ with $t\in\IC^*$. Here and in the
following, $\a,\b,\ldots=1,2$ and $\ad,\bd,\ldots=1,2$ are spinor indices. 
Consider now the space $\CP^3=\IC P^3\setminus\IC P^1$ in which the
$\l_\ad$ are not simultaneously zero. 
This space can be covered by two coordinate patches,
say $\CU_+$ ($\l_{\dot1}\neq0$) and $\CU_-$ ($\l_{\dot2}\neq0$),
with coordinates
\eqn\coord{\eqalign{
           z^\a_+\ =\ {\o^\a\over\l_{\dot1}},\qquad
           z^3_+\ =\ {\l_{\dot2}\over\l_{\dot1}}\ =:\ \l_+
           \qquad&{\rm on}\qquad\CU_+\ ,\cr
           z^\a_-\ =\ {\o^\a\over\l_{\dot2}},\qquad
           z^3_-\ =\ {\l_{\dot1}\over\l_{\dot2}}\ =:\ \l_-
           \qquad&{\rm on}\qquad\CU_-\ .\cr
}}
On the intersection $\CU_+\cap\CU_-$, they are related by
\eqn\coordrel{z_+^\a\ =\ \l_+ z_-^\a\qquad{\rm and}\qquad \l_+\ =\ \l_-^{-1}\ .
              } 
It follows from \coord\ and \coordrel\ that $\CP^3$ coincides with the total 
space of the rank $2$ holomorphic vector bundle\foot{Recall that the 
holomorphic line bundle $\CO(m)$ over $\IC P^1$ is defined by the transition
function $\l_+^m$.} $\IC^2\otimes\CO(1)$ over the
Riemann sphere $\IC P^1$. We denote the covering of the latter by 
$U_\pm=\CU_\pm\cap\IC P^1$ with coordinates $\l_\pm$ on $U_\pm$.

\smallskip
{\bf Real structure.}
As we are eventually interested in real fields, we need to introduce a real
structure on $\CP^3$. This can be done by defining the anti-linear 
transformations 
\eqn\realstr{\left(\matrix{\o^1\cr\o^2\cr}\right)\ \mapsto\ 
             \left(\matrix{-{\bar\o}^2\cr{\bar\o}^1\cr}\right)
             \qquad{\rm and}\qquad
              \left(\matrix{\l_{\dot1}\cr\l_{\dot2}\cr}\right)\ \mapsto\ 
             \left(\matrix{-{\bar\l}_{\dot2}\cr{\bar\l}_{\dot1}\cr}\right)
}
which induce an antiholomorphic involution (real structure) $\tau$ on $\CP^3$.
On the coordinates \coord, it acts as follows:
\eqn\tauaction{\tau(z^1_\pm,z^2_\pm,z^3_\pm)\ =\ \left(
                \pm{{\bar z}^2_\pm\over{\bar z}^3_\pm},
                \mp{{\bar z}^1_\pm\over{\bar z}^3_\pm},
                -{1\over{\bar z}^3_\pm}\right)\ .}
It is not difficult to see that $\tau$ has no fixed points in $\CP^3$ but does 
leave invariant projective lines $\IC P^1$ joining the points $p$ and $\tau(p)$
for any $p\in\CP^3$. For two other possible real structures on $\CP^3$, see 
e.g. \PopovRB.

\smallskip
{\bf Real rational curves.}
Holomorphic sections of the vector bundle $\CP^3\to\IC P^1$ are rational degree
one curves $\IC P^1_x\hookrightarrow\CP^3$ defined by the equations
\eqn\twistoreq{z^\a_\pm\ =\ x^{\a\ad}\l^\pm_\ad\quad{\rm for}\quad
               (\l_\ad^+)\ =\ \left(\matrix{1\cr\l_+}\right)\quad{\rm and}
               \quad\l_\ad^-\ =\ \l_+^{-1}\l_\ad^+\quad{\rm with}\quad
               \l_\pm\in U_\pm}
and parametrized by the moduli $(x^{\a\ad})\in\IC^4$. Thus, the
complexified spacetime $\IC^4$ can be identified with the moduli space of
holomorphic sections of the bundle $\CP^3\to\IC P^1$. 

The action of the map $\tau$ on the $x^{\a\ad}$ is given by
\eqn\actiontaui{\tau\left(\matrix{x^{1{\dot1}} & x^{1{\dot2}}\cr
                                  x^{2{\dot1}} & x^{2{\dot2}}\cr}
                    \right)\ =\ 
                    \left(\matrix{{\bar x}^{2{\dot2}} & 
                                  -{\bar x}^{2{\dot1}}\cr
                                  -{\bar x}^{1{\dot2}} & 
                                  {\bar x}^{1{\dot1}}\cr}\right)\ ,}
where the overbar denotes complex conjugation. The real subspace $\IR^4$
of $\IC^4$ invariant under $\tau$ is defined by the equations
\eqn\actiontauii{x^{2{\dot2}}\ =\ {\bar x}^{1{\dot1}}\ =:\ x^1-\i x^2
                 \qquad{\rm and}\qquad
                 x^{2{\dot1}}\ =\ -{\bar x}^{1{\dot2}}\ =:\ x^3-\i x^4
}
and parametrized by the real coordinates $x=(x^\mu)\in\IR^4$ with 
$\mu=1,\ldots,4$. The action of $\tau$ on arbitrary functions 
$f(x,\l_\pm,{\bar\l}_\pm)$ is then immediate. Hence, one may 
impose the condition of invariance under $\tau$ on sections of the form
\twistoreq\ and realize that they are real if the conditions \actiontauii\ 
hold true. 

\smallskip
{\bf Metric and signature.}
On the space $\IR^4$ of $\tau$-real holomorphic curves 
$\IC P^1_x\hookrightarrow\CP^3$, one can introduce the metric
\eqn\metric{{\rm d}s^2\ =\ \det({\rm d}x^{\a\ad})\ =\ g_{\mu\nu}{\rm d}x^\mu
            {\rm d}x^\nu\ ,}
which is Euclidean, $(g_{\mu\nu})={\rm diag}(+1,+1,+1,+1)$, due to 
\actiontauii. Other real structures on $\CP^3$ produce a metric of
signature $({+}\,{+}\,{-}\,{-})$ on $\IR^4$ (see e.g. \PopovRB).

Note that topologically, $\CP^3$ is the direct product 
\eqn\realts{\IR^4\times\IC P^1\ \cong\ \CP^3}
with coordinates $(x^\mu,\l_\pm)$.
Therefore, on the patches $\CU_\pm$ covering $\CP^3$ one can use either 
the complex coordinates \coord\ or coordinates $(x^\mu,\l_\pm)$. 
In fact, the formulas \twistoreq\ together with their complex conjugates define
a diffeomorphism of $\IR^4\times\IC P^1$ onto $\CP^3$. Its inverse is given
by
\eqn\invtrafo{x^{1{\dot1}}\ =\ {z^1_++{\bar z}^3_+{\bar z}^2_+\over
                                1+z^3_+{\bar z}^3_+}\ =\
                               {{\bar z}^3_-z^1_-+{\bar z}^2_-\over
                               1+z^3_-{\bar z}^3_-}\ ,\quad
              x^{1{\dot2}}\ =\ -{z^2_+-{\bar z}^3_+{\bar z}^1_+\over
                                1+z^3_+{\bar z}^3_+}\ =\
                               -{{\bar z}^3_-z^2_--{\bar z}^1_-\over
                               1+z^3_-{\bar z}^3_-}\ ,}
together with $\l_\pm=z^3_\pm$ and \actiontauii\ for $x^{2{\dot2}}$ and
$x^{2{\dot1}}$.

\smallskip
{\bf Vector fields.}
On the complex manifold $\CP^3$, we have a natural basis 
$(\partial/\partial{\bar z}^\a_\pm,\partial/\partial{\bar z}^3_\pm)$ for
antiholomorphic vector fields. These 
are sections of the bundles\foot{Here, $\cob{m}$ denotes the bundle which is 
complex conjugate to $\CO(m)$.} $\cob{-1}$ and
$\cob{-2}$, respectively. Using the formulas \invtrafo, we can express them 
in terms of the coordinates $(x^{\a\ad},\l_\pm)$ according to
\eqn\vectfield{\eqalign{
               &\partial_{{\bar z}^1_\pm}\ =\ -\g_\pm\l^\ad_\pm\partial_{2\ad}\
               =:\ -\g_\pm{\bar V}^\pm_2\ ,\qquad
               \partial_{{\bar z}^2_\pm}\ =\ \g_\pm\l^\ad_\pm\partial_{1\ad}\
               =:\ \g_\pm{\bar V}^\pm_1\ ,\cr
                &\kern1.5cm\partial_{{\bar z}^3_\pm}\ =\ 
                \partial_{{\bar\l}_\pm}+
                \g_\pm x^{\a{\dot1}}{\bar V}^\pm_\a\ =:\
                {\bar V}^\pm_3+\g_\pm x^{\a{\dot1}}{\bar V}^\pm_\a\ ,
}}
where 
\eqn\vectfielda{\l_\pm^\ad\ =\ \epsilon^{\ad\bd}\l^\pm_\bd\qquad{\rm with}
                \qquad \epsilon^{{\dot1}{\dot2}}\ =\ 1
              \qquad{\rm and}\qquad \g_\pm\ =\ (1+\l_\pm{\bar\l}_\pm)^{-1}\ .} 
We use $\epsilon_{\ad\bd}\epsilon^{\bd\gd}=\d^\gd_\ad$. Thus, the vector fields
\eqn\vectfieldi{{\bar V}^\pm_\a\ =\ 
                  \l^\ad_\pm\partial_{\a\ad}\qquad{\rm and}
                \qquad {\bar V}^\pm_3\ =\ \partial_{{\bar\l}_\pm}}
form a basis of vector fields of type $(0,1)$ on $\CU_\pm\subset\CP^3$ in the
coordinates $(x^{\a\ad},\l_\pm)$. For more details on the geometry of the 
twistor space $\CP^3$, see e.g. \PopovPC\ and references therein.

\newsec{Holomorphic Chern-Simons theory on  $W\IC P^{3|2}(1,1,1,1|p,q)$} 

{\bf Weighted projective space $W^{3|2}(p,q)$.}
Let us consider the weighted projective space 
$W^{3|2}(p,q)\!=\!W\IC P^{3|2}(1,1,1,1|p,q)$ with four homogeneous bosonic 
coordinates $(\o^\a,\l_\ad)$ of weight one and two fermionic coordinates
$(\rho_i)=(\rho_1,\rho_2)$ of weight $p$ and $q$, respectively. Here, 
$(\o^\a,\l_\ad)$ are the homogeneous coordinates on $\IC P^3$. Together, the
coordinates $(\o^\a,\l_\ad,\rho_i)$ are subject to the equivalence relation
$(\o^\a,\l_\ad,\rho_1,\rho_2)\sim(t\o^\a,t\l_\ad,t^p\rho_1,t^q\rho_2)$ for
all $t\in\IC^*$. We assume that the sum of fermionic weights is equal to the 
sum of bosonic weights, i.e. $p+q=4$. In this case,
 $W^{3|2}(p,q)$ is a CY supermanifold and one may consider the topological
$B$-model of $W^{3|2}(p,q)$ \WittenNN. The same arguments
as given in \WittenNN\ then yield hCS theory on this space.

\smallskip
{\bf Action of hCS theory.}
Let $\CE$ be a trivial rank $n$ complex vector bundle
over  $W^{3|2}(p,q)$ and $\CA$ a connection one-form on $\CE$. 
Consider a subspace $\CY$ of  $W^{3|2}(p,q)$ which is parametrized by the
holomorphic and antiholomorphic bosonic coordinates as well as the holomorphic
fermionic coordinates.\foot{Recall that $\CY$ is the world-volume of the 
$D$-branes which are not quite space-filling \WittenNN.} 
The action of hCS theory then reads as \WittenNN
\eqn\hcs{S_{\rm hCS}\ =\ \int_\CY\,\Omega\wedge{\rm tr}\left(\CA^{0,1}\wedge
         {\bar\partial}\CA^{0,1}+{2\over3}\CA^{0,1}\wedge\CA^{0,1}\wedge
         \CA^{0,1}\right)\ ,}
where $\Omega$ is a nowhere vanishing holomorphic volume form and
$\CA^{0,1}$ denotes the $(0,1)$-component of $\CA$. It is assumed that 
$\CA^{0,1}$ contains no antiholomorphic fermionic components and depends only
on $\rho_i$ as well as on $(\o^\a,\l_\ad)$ and $({\bar\o}^\a,{\bar\l}_\ad)$.
The action \hcs\ leads to the field equations
\eqn\hcseom{{\bar\partial}\CA^{0,1}+\CA^{0,1}\wedge\CA^{0,1}\ =\ 0\ .}
Note that for constructing solutions to these equations, one can use
a generalization of the Riemann-Hilbert problem based on the equivalence of the
\v Cech and Dolbeault descriptions of holomorphic bundles \PopovFB.

\smallskip
{\bf Local coordinates.}
Consider now an open subset of  $W^{3|2}(p,q)$ defined as
$\CW\CP^{3|2}(p,q)= W\IC P^{3|2}(1,1,1,1|p,q)\setminus W\IC P^{1|2}(1,1|p,q)$.
It can be covered by two patches, which we denote by $\hCU_\pm$. Since the 
body of the CY supermanifold $\CW\CP^{3|2}(p,q)$ is the twistor space $\CP^3$,
we may take the coordinates given in \coord\ as bosonic coordinates on 
$\hCU_\pm$ and as fermionic ones
\eqn\coordi{\rho_1^+\ =\ {\rho_1\over\l_{\dot1}^p},\quad
            \rho_2^+\ =\ {\rho_2\over\l_{\dot1}^q}
           \quad {\rm on}\quad\hCU_+\quad{\rm and}\quad
           \rho_1^-\ =\ {\rho_1\over\l_{\dot2}^p},\quad
           \rho_2^-\ =\ {\rho_2\over\l_{\dot2}^q}
           \quad{\rm on}\quad\hCU_-\ ,}
which are related by $\rho_1^+=\l_+^p\rho_1^-$ and $\rho_2^+=\l_+^q\rho_2^-$ 
on the intersection $\hCU_+\cap\hCU_-$. Note that as bosonic coordinates on
$\CW\CP^{3|2}(p,q)$, we can use either $(z^\a_\pm,z_\pm^3)$ or 
$(x^{\a\ad},\l_\pm)$ and similarly for the vector fields \vectfield\
and \vectfieldi. The vector fields \vectfieldi\ together with 
$\partial/\partial{\bar\rho}_i$ form a basis of vector fields of type $(0,1)$
on $\CW\CP^{3|2}(p,q)$.

\smallskip
{\bf HCS theory on $\CW\CP^{3|2}(p,q)$.}
Having given all the ingredients, we may now consider hCS theory on 
$\CW\CP^{3|2}$ with $p+q=4$. The equations of motion \hcseom\ on the patches 
$\hCU_\pm$ of $\CW\CP^{3|2}$ read 
\eqn\hcseomII{\eqalign{
              {\bar V}^\pm_\a\CA_\b^\pm-{\bar V}^\pm_\b\CA_\a^\pm
              +[\CA_\a^\pm,\CA_\b^\pm]\ &=\ 0\ ,\cr
               {\bar V}^\pm_\a\CA_{{\bar\l}_\pm}-\partial_{{\bar\l}_\pm}
               \CA_\a^\pm+[\CA_\a^\pm,\CA_{{\bar\l}_\pm}]\ &=\ 0\ ,\cr
}}
where 
\eqn\compofa{
      \CA_\a^\pm\ :=\  {\bar V}^\pm_\a\lrcorner\CA^{0,1}\qquad{\rm and}\qquad
     \CA_{{\bar\l}_\pm}\ :=\  \partial_{{\bar\l}_\pm}\lrcorner\CA^{0,1}\ .}
Here, ``$\lrcorner$'' denotes the interior product of vector fields with
differential forms. As usual in the twistor approach, we shall assume 
that there exists a gauge in which the components $ \CA_{{\bar\l}_\pm}$ are 
zero. This corresponds to the holomorphic triviality on any
$\IC P^1_{x,\rho}\hookrightarrow\CW\CP^{3|2}(p,q)$
of the holomorphic vector bundle which is  associated
to any solution of hCS theory.

Note that generically, the gauge field $\CA^{0,1}$ on $\CW\CP^{3|2}(p,q)$
can be expanded in terms of the odd coordinates $\rho_i$ according to
\eqn\fieldexpageneric{
      \CA^{0,1}\ =\ A+\rho_i\psi^i+\rho_1\rho_2 G\ ,
}
Linearizing the equations \hcseom\ around
the trivial solution $\CA^{0,1}=0$, we see that ${\bar\partial}\CA^{0,1}=0$
or equivalently ${\bar\partial}A=0$,  ${\bar\partial}\psi^i=0$ and
${\bar\partial}G=0$. Since $\CA^{0,1}$ is $\CO(0)$-valued
and $\rho_1$ and $\rho_2$ take values in $\CO(p)$ and $\CO(q)$, respectively,
the fields $A$, $\psi^1$, $\psi^2$ and $G$ are $(0,1)$-forms on $\CP^3$ 
with values in the bundles $\CO(0)$, $\CO(-p)$, $\CO(-q)$ and $\CO(-4)$.  
Therefore, each of these fields 
determines an element of the sheaf cohomology group $H^1(\CP^3,\CO(2s-2))$,
where $s=1$ for $A$, $s=1-{1\over2}p$ for  $\psi^1$,  
$s=1-{1\over2}q$ for  $\psi^2$ and $s=-1$ for $G$.
According to the Penrose transform \PenroseIN, elements of these cohomology 
groups 
correspond to solutions of the field equations for massless fields of
helicity $s$ on the space $\IR^4$; the latter being parametrized by 
$x^{\a\ad}$. In the next section, we describe this correspondence
beyond linearized level.

\newsec{Field models on $\IR^4$ equivalent to hCS theory on 
        $\CW\CP^{3|2}(p,q)$}

{}From now on, we shall work in the patch 
$\hCU_+$ of $\CW\CP^{3|2}(p,q)$ and for notational simplicity we omit the
subscript and the superscript ``$+$'' in all expressions except when confusion
may arise.

{\bf HCS theory on  $\CW\CP^{3|2}(1,3)$.} 
Let us consider the case $p=1$ and $q=3$, where the fermionic coordinates 
$\rho_1$ and $\rho_2$ take values in the bundles $\CO(1)$ and $\CO(3)$, 
respectively. From \vectfieldi\ and \compofa\ one
concludes that the components $\CA_1$, $\CA_2$ and $\CA_{\bar\l}$ of the
$(0,1)$-form $\CA^{0,1}$ take values in the bundles $\CO(1)$, $\CO(1)$ and
$\cob{-2}$. This fixes the dependence of $\CA_\a$ and $\CA_{\bar\l}$ on
$\l$ and ${\bar\l}$ up to gauge transformations. Namely, we obtain
\eqn\fieldexpa{\eqalign{
      \CA_\a\ &=\ \l^\ad A_{\a\ad}+\rho_1\chi_\a+{1\over2!{\sqrt3}}
           \rho_2\g^2\hl^\ad\hl^\bd
      {\widetilde\chi}_{\a\ad\bd}+{1\over3!}\rho_1\rho_2\g^3
            \hl^\ad\hl^\bd\hl^\gd
      G_{\a\ad\bd\gd}\ ,\cr
      \CA_{\bar\l}\ &=\ {1\over{\sqrt3}}
              \rho_2\g^3\hl^\ad{\widetilde\chi}_\ad+{1\over2!}
      \rho_1\rho_2\g^4\hl^\ad\hl^\bd G_{\ad\bd}\ ,
}}
where 
\eqn\defofth{\hl^\ad\ :=\ \tau(\l^\ad)\ =\ {\bar\l}_\ad\ .} 
In \fieldexpa, all the fields $A_{\a\ad}$, $\chi_\a$, ${\widetilde\chi}_\ad$,
$G_{\ad\bd}$, ${\widetilde\chi}_{\a\ad\bd}$ and $G_{\a\ad\bd\gd}$
depend only on $(x^{\a\ad})\in\IR^4$. 

Substituting \fieldexpa\ into \hcseomII, we get the equations
\eqn\condiI{{\widetilde\chi}_{\a\ad\bd}\ =\ -\nabla_{\a(\ad}
            {\widetilde\chi}_{\bd)}\qquad{\rm and}\qquad
            G_{\a\ad\bd\gd}\ =\ -\nabla_{\a(\ad}G_{\bd\gd)}}
showing that ${\widetilde\chi}_{\a\ad\bd}$ and $G_{\a\ad\bd\gd}$ are 
composite fields describing no 
independent degrees of freedom. Here, parentheses denote normalized 
symmetrization with respect to the enclosed indices and $\nabla_{\a\ad}=
\partial_{\a\ad}+[A_{\a\ad},\ \cdot\ ]$.  The remaining equations are given by
\eqna\eomI
$$\eqalignno{
       f_{\ad\bd}\ =\ -{\sfrac{1}{2}}
              \epsilon^{\a\b}(\partial_{\a\ad}A_{\b\bd}-
         \partial_{\b\bd}A_{\a\ad}+[A_{\a\ad},A_{\b\bd}])\ &=\ 0\ , &\eomI a\cr
          \nabla_{\a\ad}\chi^\a\ &=\ 0\ ,&\eomI b\cr
          \nabla_{\a\ad}{\widetilde\chi}^\ad\ &=\ 0\ ,&\eomI c\cr
          \epsilon^{\ad\bd}\nabla_{\a\ad}G_{\bd\gd}-\{\chi_\a,
          {\widetilde\chi}_\gd\}\ &=\ 0\ .&\eomI d
}$$
Due to \eomI{}  and the Bianchi identities, all other equations following from 
\hcseomII\ are automatically satisfied. 

The equations \eomI{} are the equations of motion for the action functional 
\eqn\actionI{S\ =\ \int\,{\rm d}^4x\ {\rm tr}\,\left\{
                    G^{\ad\bd}f_{\ad\bd}+
                    {\widetilde\chi}^\ad\nabla_{\a\ad}\chi^\a
                 \right\}}
which can be obtained from \hcs\ by integration over the fermionic coordinates 
and over the sphere $\IC P^1$. Note that this action has an obvious 
supersymmetry. The transformation laws are given by
\eqn\supersymm{\eqalign{
         \d_\xi A_{\a\ad}\ =\ \xi_\ad\chi_\a\qquad&{\rm and}\qquad
  \d_\xi G_{\ad\bd}\ =\ -\epsilon^{\a\b}\xi_{(\ad}\nabla_{\a\bd)}\chi_\b\ ,\cr
            \qquad\d_\xi \chi_\a\ =\ 0\qquad&{\rm and}\qquad
           \d_\xi{\widetilde\chi}_\ad\ =\ -\xi^\bd(G_{\ad\bd}+f_{\ad\bd})\ ,\cr
}}
where $\xi_\ad$ is a constant (anticommuting) spinor.
The action \actionI\ describes a truncation of the self-dual
$\CN=4$ SYM model \SiegelZA\ for which all the scalars and three of 
the dotted and three of the undotted fermions are put to zero. 

\smallskip
{\bf HCS theory on $\CW\CP^{3|2}(2,2)$.}
Now we consider the case $p=q=2$, i.e. the fermionic coordinates $\rho_i$
take values in the line bundle $\CO(2)$. The  equations of motion of hCS theory
on  $\CW\CP^{3|2}(2,2)$ have the same form \hcseomII. Again, the functional
dependence on $\l$ and ${\bar\l}$ is fixed up to gauge transformations by the 
geometry of $\CW\CP^{3|2}(2,2)$. Namely, this dependence has the form
\eqn\fieldexpaII{\eqalign{
      \CA_\a\ &=\ \l^\ad A_{\a\ad}+{1\over2!}
                  \rho_i\g\hl^\ad\phi^i_{\a\ad}+
                  {1\over3!}\rho_1\rho_2\g^3\hl^\ad\hl^\bd\hl^\gd
                  G_{\a\ad\bd\gd}\ ,\cr
      \CA_{\bar\l}\ &=\ {1\over 2!}\rho_i\g^2\phi^i+{1\over2!}\rho_1\rho_2\g^4
                    \hl^\ad\hl^\bd G_{\ad\bd}\ ,
}}
where again the coefficient fields do only depend on $(x^{\a\ad})\in\IR^4$
and $\l^\ad$, $\g$ and $\hl^\ad$ are given in \vectfielda\ and \defofth,
respectively.

Substituting
\fieldexpaII\ into \hcseomII, we obtain the following equations:
\eqn\condiII{\phi^i_{\a\ad}\ =\ -\nabla_{\a\ad}\phi^i
            \qquad{\rm and}\qquad
            G_{\a\ad\bd\gd}\ =\ -\nabla_{\a(\ad}G_{\bd\gd)}\ .}
The remaining nontrivial equations read 
\eqna\eomII
$$\eqalignno{f_{\ad\bd}\ &=\ 0\ ,&\eomII a\cr
          \nabla_{\a\ad}\nabla^{\a\ad}\phi^i\ &=\ 0\ ,&\eomII b\cr
          \epsilon^{\ad\bd}\nabla_{\a\ad}G_{\bd\gd}-{\sfrac{3}{4}}
          \epsilon_{ij}\{\phi^i,\nabla_{\a\gd}\phi^j\}\ &=\ 0\ .&\eomII c\cr
}$$
The associated action functional is given by
\eqn\actionII{S\ =\ \int\,{\rm d}^4x\ {\rm tr}\,\left\{
                    G^{\ad\bd}f_{\ad\bd}+{\sfrac{3}{8}}
                     \epsilon_{ij}\phi^i\nabla_{\a\ad}
                     \nabla^{\a\ad}\phi^j\right\}\ ,}
and can be obtained from \hcs. Note that formally \actionII\ looks as the 
bosonic truncation of the self-dual $\CN=4$ SYM theory \SiegelZA, i.e.~all the
spinors and four of the six scalars of self-dual $\CN=4$ super SYM theory
are put to zero. However, in \actionII\ the parity of the scalars $\phi^i$ is 
different, as they are Gra{\ss}mann odd. To understand their nature, note that
in the expansions \fieldexpaII\ we have two Gra{\ss}mann odd vectors 
$\phi^i_{\a\ad}$ which satisfy the equations
\eqna\eqforsca
$$\eqalignno{\epsilon^{\a\b}\nabla_{\a\ad}\phi^i_{\b\bd}\ &=\ 0\ ,
              &\eqforsca a\cr
             \qquad \nabla^{\a\ad}\phi^i_{\a\ad}\ &=\ 0
              &\eqforsca b\cr 
}$$
following from \hcseomII. Solutions to these equations describe tangent vectors
$\d A_{\a\ad}$ (with assigned odd parity) to the solution space of the 
self-duality equations \eomII{a} \refs{\WittenZE,\VafaTF}. However, due to the
first equation in \condiII\ (which solves \eqforsca{a} and reduces 
\eqforsca{b} to \eomII{b}), the $\phi^i_{\a\ad}$ are projected to zero in the
moduli space of solutions to the equations \eomII{a}. By choosing 
$\phi^i=0$, we remain with the equations 
\eqn\siegel{f_{\ad\bd}\ =\ 0\qquad{\rm and}\qquad
            \nabla_{\a\ad}G^{\ad\bd}\ =\ 0\ ,} 
which can be obtained from the Lorentz invariant Siegel action \SiegelZA
\eqn\actionIV{S\ =\ \int\,{\rm d}^4x\ {\rm tr}\ \{ G^{\ad\bd}f_{\ad\bd}\}\ ,}
describing purely bosonic self-dual Yang-Mills theory.

\smallskip
{\bf HCS theory on $\CW\CP^{3|2}(4,0)$.}
Finally, we want to discuss the case in which the fermionic coordinate $\rho_1$
has weight four and $\rho_2$ weight zero, i.e. we consider  
$\CW\CP^{3|2}(4,0)$. The field equations of hCS theory on $\CW\CP^{3|2}(4,0)$ 
have the form \hcseomII, where the vector fields ${\bar V}_\a^\pm$ and 
$\partial_{{\bar\l}_\pm}$ are given in \vectfieldi. Proceeding as in the 
previous two subsections, we obtain the following field expansions:
\eqn\fieldexpaIII{\eqalign{
      \CA_\a\ &=\ \l^\ad A_{\a\ad}+
                  {1\over3!}\rho_1\g^3\hl^\ad\hl^\bd\hl^\gd
                  \chi_{\a\ad\bd\gd}+\rho_2\l^\ad\psi_{\a\ad}+
                  {1\over3!}\rho_1\rho_2
                  \g^3\hl^\ad\hl^\bd\hl^\gd G_{\a\ad\bd\gd}\ ,\cr
      \CA_{\bar\l}\ &=\ {1\over 2!}\rho_1\g^4\hl^\ad\hl^\bd 
                   \chi_{\ad\bd}+{1\over 2!}\rho_1\rho_2\g^4\hl^\ad\hl^\bd 
                  G_{\ad\bd}\ .
}}
All the Lie-algebra valued fields appearing in the expansions \fieldexpaIII\
depend only on $(x^{\a\ad})\in\IR^4$. Note also that $A_{\a\ad}$, $G_{\ad\bd}$
and $G_{\a\ad\bd\gd}$ are bosonic while $\psi_{\a\ad}$, $\chi_{\ad\bd}$ and
$\chi_{\a\ad\bd\gd}$ are fermionic fields.

Substituting \fieldexpaIII\ into \hcseomII, we obtain
\eqn\condiIII{\chi_{\a\ad\bd\gd}\ =\ -\nabla_{\a(\ad}\chi_{\bd\gd)}
              \qquad{\rm and}\qquad
              G_{\a\ad\bd\gd}\ =\ -\nabla_{\a(\ad}G_{\bd\gd)}
              -\{\psi_{\a(\ad},\chi_{\bd\gd)}\}\ .}
The remaining nontrivial equations are
\eqna\eomIII
$$\eqalignno{f_{\ad\bd}\ &=\ 0\ , & \eomIII a\cr
             \epsilon^{\a\b}\nabla_{\a(\ad}\psi_{\b\bd)}\ &=\ 0\ , 
             & \eomIII b\cr
             \nabla_{\a\ad}G^{\ad\bd}+\{\psi_{\a\ad},\chi^{\ad\bd}\}\ &=\ 0\
             , & \eomIII c\cr
             \nabla_{\a\ad}\chi^{\ad\bd}\ &=\ 0\ . & \eomIII d\cr
}$$
In this case, the action functional from which these equations arise is
\eqn\actionIII{S\ =\ \int\,{\rm d}^4x\ {\rm tr}\,\left\{
                    G^{\ad\bd}f_{\ad\bd}+\epsilon^{\a\b}
                     (\nabla_{\a\ad}\psi_{\b\bd})\chi^{\ad\bd}\right\}\ .}
This time, the multiplet contains a spacetime vector 
$\psi_{\a\ad}$ and an anti-self-dual two-form $\chi_{\ad\bd}$ which
are both Gra{\ss}mann odd. Such fields are well known from topological 
Yang-Mills theories \refs{\WittenZE,\VafaTF}. In this respect, the model
\eomIII{}, \actionIII\ can be understood as a truncated self-dual sector of 
these theories.\foot{One may also consider more than two fermionic coordinates
in order to enlarge the multiplet. This may lead to other truncations of
topological Yang-Mills theories.}

\bigbreak\bigskip\centerline{{\bf Acknowledgments}}\nobreak \smallskip

We thank Olaf Lechtenfeld, Christian S\"amann and Robert Wimmer for useful
comments. This work was partially supported by the Deutsche 
Forschungsgemeinschaft (DFG).

\appendix{A}{Signature $({+}\,{+}\,{-}\,{-})$}

{}For the Kleinian space $\IR^{2,2}=(\IR^4,g)$ with the metric 
$g={\rm diag}(+1,+1,-1,-1)$, all the calculations and formulas are similar. 
What should be changed is the following. Instead of the reality conditions
\actiontauii, one considers
\eqn\actiontauiia{x^{2{\dot2}}\ =\ {\bar x}^{1{\dot1}}\ =:\ x^1-\i x^2
                 \qquad{\rm and}\qquad
                 x^{2{\dot1}}\ =\ {\bar x}^{1{\dot2}}\ =:\ x^3+\i x^4
}
which correspond to the involution (real structure)
\eqn\tauactionA{{\hat\tau}(z^1_\pm,z^2_\pm,z^3_\pm)\ =\ \left(
                {{\bar z}^2_\pm\over{\bar z}^3_\pm},
                {{\bar z}^1_\pm\over{\bar z}^3_\pm},
                {1\over{\bar z}^3_\pm}\right)}
on the twistor space $\CP^3$. Furthermore, instead of 
$\CP^3$ one should consider the space\foot{For more details see e.g.
\LechtenfeldIK.}
${\widetilde\CP}^3=\CP^3\setminus\CT^3$, where $\CT^3$ is the fixed point
set under the involution ${\hat\tau}$. In fact, $\CT^3$ is the
three-dimensional real manifold $\IR P^3\setminus\IR P^1$ fibered over 
$S^1\cong\IR P^1\subset\IC P^1$. All CY supermanifolds $\CW\CP^{3|2}(p,q)$
considered above are now supermanifolds with ${\widetilde\CP}^3$
as body. One should also substitute $\IC P^1$ by the
two-sheeted hyperboloid $H^2=\IC P^1\setminus S^1=H_+\cup H_-$ and use
\eqn\last{(\hl_+^\ad)\ =\ \left(\matrix{-1\cr{\bar\l}_+\cr}\right),\quad
          (\hl_-^\ad)\ =\ \left(\matrix{-{\bar\l}_-\cr1\cr}\right),\quad
          \g_\pm=\pm(1-\l_\pm{\bar\l}_\pm)^{-1}\quad{\rm with}\quad
          \l_\pm\in H_\pm}
instead of $\g_\pm$ and $\hl^\ad_\pm$ as given in \vectfielda\ and \defofth,
respectively. All other formulas, including the equations of motion for hCS 
theory and the field expansions of $\CA^{0,1}$ keep their form. The resulting 
field equations on $\IR^{2,2}$ will only differ by some signs in front of the 
interaction terms.

\listrefs
\end